\documentclass[sigconf]{acmart}

\usepackage{subcaption}
\usepackage{mathtools}
\usepackage[capitalize,noabbrev]{cleveref}

\setcopyright{none}
\acmDOI{}
\acmISBN{}
\acmConference[USRW '26]{First Unified Search and Recommendation Workshop}%
  {October 2, 2026}{Minneapolis, Minnesota, USA}
\acmYear{2026}

\begin{document}

\title{Orthogonal Norm Growth in Two-Tower Retrieval: A Popularity Mechanism for Unified Discovery}

\author{Andrei Atamaniuk}
\affiliation{%
  \institution{Wildberries and Russ}
  \country{Russia}
}
\email{atamanyuk.andrey@rwb.ru}

\begin{abstract}
Unified search and recommendation systems may update the same item vectors
from both interaction streams. We study how repeated updates can increase an
item vector's norm and affect retrieval under an unnormalized dot product. A
cosine-only loss makes the output gradient perpendicular to the current item
vector---tangent to its constant-norm sphere. Training changes encoder
parameters, however, so the realized item-vector movement need not remain
tangent. We derive sufficient conditions under which it does; every nonzero
update then increases squared norm by its squared length. When the tasks share
an item encoder, sequential updates from both can accumulate in the same norm,
which raw-inner-product serving can expose for either task. This conditional
result is not a theorem about joint SGD.

Controlled ablations test these conditions. On MovieLens~32M and Amazon Books,
accumulated tangent movement accompanies frequency--norm association and
additional head-item exposure under raw-dot rather than normalized-item
ranking. In a nonlinear tower combining a per-item ID vector with a shared
content encoding, active item movement remains near-tangent. We separate its
norm change into the ID-update contribution and other output changes, then use
a matched training intervention that removes immediate norm growth caused by
the ID-row update. The intervention reduces norm-related exposure but also
recommendation utility, so it is diagnostic rather than a proposed mitigation.
Our logs are recommendation-only: the formulation supplies a
shared-representation audit, while prevalence in joint traffic remains open.
\end{abstract}

\keywords{unified search and recommendation, neural retrieval, two-tower models,
embedding norms, popularity bias, representation geometry}

\maketitle

\section{Introduction}
\label{sec:introduction}

Search and recommendation are complementary routes to a shared catalog:
search starts from an explicit query, whereas recommendation infers an
information need from a user and context. Unified systems connect their
behavior sequences, supervision, model parameters, or item representations
\cite{yao2021user,sun2023kuaisar,liu2024usr}. At candidate-generation scale,
both tasks commonly use bi-encoder or two-tower retrieval: a request-side
encoder produces a query or user vector, an item-side encoder produces a
catalog vector, and a simple similarity enables precomputation and approximate
nearest-neighbor search \cite{huang2013dssm,huang2024retrievalsurvey}. Sharing
the item tower is efficient, but it also lets training on one surface change
the representation served on the other. Consider an item that receives
repeated search-side updates. Under the conditions derived below, those
updates add to the squared norm of the same catalog vector later used by
recommendation. If recommendation retrieves with raw inner products, this
search-derived magnitude enters its scores even if its request tower is
unchanged; normalizing item vectors before serving removes this particular
channel. This cross-surface dependency motivates measuring norm growth
separately by task.

We study the norm component of this cross-surface dependency. If item
frequency becomes correlated with item-vector norm, raw-inner-product ranking
contains a frequency-associated component:
\begin{equation}
  \langle u,q_i\rangle
  = \|u\|\,\|q_i\|\,\cos(u,q_i).
  \label{eq:score-decomposition}
\end{equation}
That component may improve utility, amplify exposure, or do both; popularity is
not intrinsically harmful
\cite{klimashevskaia2024popularitybias,tsyplov2025compound}. Our question
is mechanistic: \emph{when can repeated updates under an objective whose
item-output dependence is only through cosine similarities accumulate item
norm, how can two tasks contribute to the same norm, and when does that norm
affect retrieval exposure?}

Throughout, the \emph{output gradient} is the derivative with respect to the
current item vector $q_i$, whereas the \emph{realized displacement}
$\Delta q_i$ is its before--after change caused by an optimizer step. We call
a gradient or displacement \emph{tangent} at $q_i$ when its inner product with
$q_i$ is zero, meaning that the two vectors are perpendicular. A common
argument is that a cosine objective has a tangent output gradient, so the item
vector itself must move tangentially and increase its norm
\cite{wang2017normface,draganov2024pitfalls,draganov2025embeddingnorms}.
A deep encoder instead updates parameters; its Jacobian can rotate the output
gradient and mix it with gradients from other examples. We therefore
distinguish
\[
 q_i^\top\nabla_{q_i}\mathcal L=0
 \quad\text{from}\quad
 q_i^\top\Delta q_i=0,
\]
and identify sufficient conditions under which the first statement does imply
the second.

The theorem below gives narrow, testable sufficient conditions rather than a
description of every modern tower. Outside those conditions, tangent output
motion must be measured rather than assumed. We therefore also study a
nonlinear item tower that sums a learned per-item ID vector and a shared content
encoding. The optimizer rescales the active ID row's gradient by one scalar, so
the output change due to that row remains exactly tangent; the shared content
update supplies the residual that can break tangency. Tangency of the total
item-vector
movement is then a measurable property rather than a guarantee.

Our contributions are:
\begin{itemize}
  \item \textbf{Update-level theory.} We derive the parameter-to-output update
  and prove sufficient conditions for exact tangent item motion, showing when
  a tangent output gradient does---and does not---imply tangent movement of a
  parameterized item vector.
  \item \textbf{Shared-stream accumulation.} Under uniform interaction
  sampling and conditional on trajectory-dependent increments, sequential
  task-labeled search and recommendation updates contribute additively to one
  shared item norm. We state explicitly what this accounting does and does not
  prove about joint SGD or mixed-task minibatches.
  \item \textbf{Mechanism-oriented evidence.} On MovieLens~32M and Amazon
  Books, condition tests, total-output diagnostics, exact path accounting, and
  a matched intervention connect tangent motion to frequency-associated norm
  growth and norm-induced head exposure, while reporting the accompanying
  utility cost.
  \item \textbf{Unified-discovery audit.} We translate the mechanism into
  task-separated diagnostics for shared item towers and serving indices. Our
  experiments use recommendation logs, so search-side prevalence remains a
  scoped validation target rather than a claimed result.
\end{itemize}

\paragraph{Artifacts.}
Code, configurations, and saved outputs are in the
\href{https://gitlab.com/wrapper228/usrw_pb/-/blob/main/README.md}
{repository (click here)}. Full derivations, expanded protocols,
confidence intervals, and diagnostic tables are in the
\href{https://gitlab.com/wrapper228/usrw_pb/-/blob/main/usrw26_supplementary.pdf}
{supplementary document (click here)}. The paper contains the complete claims
and principal evidence.

\section{Positioning}
\label{sec:related-work}

\paragraph{Unified search and recommendation.}
Joint search--recommendation research has integrated heterogeneous behavior
sequences \cite{yao2021user}, introduced paired real-world logs
\cite{sun2023kuaisar}, and shared scenario representations and supervision in
industrial ranking systems \cite{liu2024usr}. These works ask how information
should transfer between tasks. We study a complementary question: what
optimizer-induced state can be transferred unintentionally when an item
representation or retrieval index is shared. The result is not a new unified
architecture; it is an audit of a coupling channel that such architectures may
contain.

\paragraph{Two-tower retrieval and item representations.}
Two-tower models support large-scale web search
\cite{huang2013dssm} and recommender candidate generation
\cite{covington2016youtube,huang2024retrievalsurvey}. A request tower may be
deep and sequential \cite{sun2019bert4rec}, while an item side can remain a
shallow ID matrix because such representations are inexpensive to precompute
and can be effective \cite{yuan2024contextgnn}. Our exact conditions apply per
tower: complexity on the request side neither establishes nor invalidates the
item-side guarantee.

\paragraph{Cosine geometry and popularity.}
Normalized objectives are known to produce output gradients tangent to the
hypersphere \cite{wang2017normface}. Recent analyses connect this fact to norm
growth \cite{draganov2024pitfalls,draganov2025embeddingnorms}. We refine the
connection by separating the output gradient from the realized output
displacement through the encoder Jacobian. In recommendation, popularity bias
has multiple data-, model-, and evaluation-level causes
\cite{klimashevskaia2024popularitybias}; representation collapse provides
another model-side explanation \cite{zhang2023dimensional}. We isolate one
frequency-associated magnitude path. Kim et al. report positive correlations
between item popularity and embedding norm, decompose inner-product scores into
angular and magnitude factors, and introduce partial or full item normalization
at inference \cite{kim2023testtime}. We therefore do not claim the association,
the algebraic score decomposition, or unit-item rescoring as novel. Our
distinct contribution is upstream: we analyze the parameter-to-output map,
derive sufficient conditions for tangent realized item motion, and account for
that motion across task-labeled updates of a shared item representation. We do
not claim this path is necessary, dominant, or harmful in every system.

\paragraph{Optimization and serving boundary.}
Contrastive objectives such as InfoNCE \cite{oord2018cpc} determine the output
gradient, but the optimizer determines how that gradient changes a
parameterized representation. Sparse recommendation systems consequently use
specialized row-wise adaptive updates \cite{ivchenko2022torchrec}, while dense
branches commonly use coordinate-wise adaptive optimizers. Our theorem covers
direction-preserving row scalings; the hybrid study makes the residual from a
dense shared branch explicit. Separately, training with normalized scores does
not force serving to discard magnitude. Large-scale two-tower infrastructure
may consume raw inner products for candidate retrieval or pass embeddings to a
later scorer \cite{huang2024retrievalsurvey}. We therefore treat raw-versus-unit
ranking as a serving decomposition, not as an assumption that every deployed
system deliberately mismatches its training score.

\section{A Shared-Tower Norm Mechanism}
\label{sec:mechanism}

\subsection{Two streams, one item representation}

We use \emph{stream} for the source of a training event and \emph{surface} for
a search or recommendation serving endpoint. Let
$a\in\{\mathrm S,\mathrm R\}$ denote the search and recommendation streams. A
task-specific request encoder produces $u^a$, while an item encoder
$q_i=q_\theta(x_i)$ is shared. Training events from stream $a$ compare $u^a$
with positive and negative item outputs through an objective whose dependence
on those outputs is only through cosine similarities.
Under uniform-over-events sampling within a stream, item $i$ has event
probability $p_i^a$. Events from the two streams may be interleaved in any
schedule; the argument below concerns the item path at each realized update.

At serving, either surface may use unit-item scoring, which excludes item
magnitude, or a raw inner product, which includes the factor $\|q_i\|$ in
Equation~\eqref{eq:score-decomposition}. Thus shared training alone does not
make norm affect rankings: a serving path must also admit magnitude.

\subsection{The missing parameter-to-output map}

Let $\mathcal I_B$ be the distinct item IDs that appear as positives or
negatives in a minibatch. For each $\ell\in\mathcal I_B$, let $q_\ell$ be its
current item-tower output, $J_\ell=\partial q_\ell/\partial\theta$, and
$\bar g_\ell$ the sum of the output-gradient contributions from all occurrences
of that item ID in the minibatch. An SGD step gives
\begin{equation}
 \begin{aligned}
 \Delta\theta
   &=-\eta\sum_{\ell\in\mathcal I_B}J_\ell^\top\bar g_\ell,\\
 \Delta q_i
   &\approx-\eta\sum_{\ell\in\mathcal I_B}
     J_i J_\ell^{\top}\bar g_\ell.
 \end{aligned}
 \label{eq:update-map}
\end{equation}
We call the output \emph{parameter-linear} when it is linear in the parameters
updated by the step; then the second relation is exact for a finite step. In a
general encoder it is only a first-order approximation. The equation exposes
two obstacles: the self-map $J_i J_i^{\top}$ may rotate $\bar g_i$, and the
cross-item maps $J_i J_\ell^{\top}$ may inject gradients belonging to other
item IDs.

We call an item encoder \emph{lookup-equivalent} when different item IDs have
independent parameter blocks and the Jacobian products satisfy
\begin{equation}
 J_i J_\ell^{\top} =
 \begin{cases}
   c_i I, & i=\ell,\\
   0, & i\ne \ell,
 \end{cases}
 \qquad c_i>0,
 \label{eq:lookup-jacobian}
\end{equation}
where $I$ is the output-space identity matrix. In plain terms, updating one
item's parameter block cannot move another item's output, and the item's own
parameter-to-output map can rescale but cannot rotate its output gradient. A
standard embedding table has $c_i=1$. A bias-free linear map applied to one-hot
item IDs has the same Jacobian property and is therefore lookup-equivalent
under this definition. By contrast, a linear map on nonorthogonal features has
$J_i J_\ell^{\top}=(x_i^\top x_\ell)I$, so cross-item terms survive. Two
consecutive linear layers are linear in their input but not jointly linear in
their parameters, so Equation~\eqref{eq:update-map} is no longer exact for a
finite step.

The vector $\bar g_i$ already aggregates repeated occurrences of item ID $i$.
We call an item-block update \emph{direction-preserving} when it multiplies the
entire current parameter-block gradient $J_i^\top\bar g_i$ by one positive
scalar shared across coordinates. Plain SGD and Exact RowWise Adagrad are
examples.
Momentum, weight decay, coordinate-wise preconditioning, shared
nonorthogonal features, or nonlinear shared parameters remove this guarantee;
they do not imply that empirical motion must be non-tangent.

\subsection{Exact tangent motion}

A loss that depends on nonzero $q$ only through cosine similarities is
invariant to positive rescaling:
$\mathcal L(\alpha q)=\mathcal L(q)$ for $\alpha>0$. Differentiating at
$\alpha=1$ gives
\begin{equation}
 q^\top\nabla_q\mathcal L=0.
 \label{eq:cosine-tangent}
\end{equation}

\begin{theorem}[Sufficient condition for tangent item motion]
\label{thm:tangent-motion}
Suppose an item output is produced by a parameter-linear,
lookup-equivalent encoder; the loss depends on that output only through cosine
similarities; and its item-specific parameter block is updated by a positive
scalar multiple of the current block gradient, without momentum or an
additional regularization term. Then the realized item update is tangent:
$q_i^\top\Delta q_i=0$.
\end{theorem}
\begin{proof}
Every occurrence of item ID $i$ is evaluated at the same pre-step output
$q_i$. Equation~\eqref{eq:cosine-tangent} therefore gives
$q_i^\top\bar g_i=0$. Lookup-equivalence removes contributions from other
item IDs. For some block-update scalar $s_i>0$, the direction-preserving rule
gives $\Delta\theta_i=-s_i J_i^\top\bar g_i$. Parameter-linearity and
Equation~\eqref{eq:lookup-jacobian} then give the exact output change
\[
 \Delta q_i=J_i\Delta\theta_i
 =-s_i J_i J_i^\top\bar g_i=-s_i c_i\bar g_i.
\]
It follows directly that $q_i^\top\Delta q_i=0$.
\end{proof}

Expanding the squared norm makes the resulting equality explicit:
\begin{equation}
 \begin{aligned}
 \|q_i^{(t+1)}\|^2-\|q_i^{(t)}\|^2
 &=2(q_i^{(t)})^\top\Delta q_i^{(t)}
   +\|\Delta q_i^{(t)}\|^2\\
 &=\|\Delta q_i^{(t)}\|^2\ge 0.
 \end{aligned}
 \label{eq:norm-increment}
\end{equation}
No large learning rate is required for this identity. Detectability depends on
cumulative tangent energy
$\sum_t\|\Delta q_i^{(t)}\|^2$ relative to initial norm variation and
non-tangent trajectory components.

\subsection{Conditional accumulation across tasks}
\label{sec:frequency-scope}

For item $i$, let $N_i^a(T)$ be its number of update events originating from
stream $a$ by time $T$, and let
$\delta_{i,k}^a=\|\Delta q_{i,k}^a\|^2$. If every interleaved item update
satisfies Theorem~\ref{thm:tangent-motion}, summing the exact per-step identity
in Equation~\eqref{eq:norm-increment} and partitioning increments by their
source stream gives the pathwise equality
\begin{equation}
 \|q_{i,T}\|^2=\|q_{i,0}\|^2+
 \sum_{k=1}^{N_i^{\mathrm S}(T)}\delta_{i,k}^{\mathrm S}+
 \sum_{k=1}^{N_i^{\mathrm R}(T)}\delta_{i,k}^{\mathrm R}.
 \label{eq:two-stream-accumulation}
\end{equation}
Condition on both nonnegative increment sequences. Coupling event draws with
common uniforms shows that the expected squared norm is nondecreasing in each
$p_i^{\mathrm S}$ and $p_i^{\mathrm R}$ separately: a higher-probability
stream contains every event of its lower-probability counterpart and possibly
more.

This accounting formalizes a shared-state channel: opportunities from either
task enter one magnitude that either raw-inner-product surface can consume. It
does \emph{not} prove that two items are ordered by popularity, that changing
one stream leaves either increment sequence fixed, or that joint training
improves or harms either task. In full SGD, frequencies affect batches,
negatives, gradients, and the shared trajectory. We therefore test below
whether opportunity count or per-update energy dominates in a less constrained
tower, and reserve search-side prevalence for joint-log validation.

Equation~\eqref{eq:two-stream-accumulation} assumes task-labeled updates are
applied sequentially. If one minibatch mixes both tasks, their aggregated
tangent components still yield a tangent total update, but its squared length
contains a cross-task inner product. It should then be recorded as one combined
increment unless the optimizer exposes separate task components; assigning that
cross term to either stream is not identified by norm accounting alone.

\section{Mechanism-Oriented Experiments}
\label{sec:experiments}

\subsection{Questions, data, and measurements}

The experiments follow the argument rather than benchmark architectures:
(Q1) do realized updates obey the predicted geometry when the sufficient
conditions hold, and how do they change when individual conditions are
violated; (Q2) does sufficient
cumulative tangent motion produce frequency--norm association and
magnitude-dependent retrieval; and (Q3) can a traceable tangent path remain
active in a nonlinear tower?

Experiments~0--7 use a hand-built graph of 25 positive interactions among 7
users and 19 items. Experiments~8--10 use MovieLens~32M
\cite{harper2015movielens} (32,000,204 interactions, 200,948 users, 84,432
interacted movies) and Amazon Books \cite{ni2019justifying}. A full-data joint
3-core of Amazon retains 35,179,026 interactions, 3,788,815 users, and
1,214,141 items. Ratings are implicit positives.

For geometry, we record outputs immediately before and after an optimizer step
and measure
$|\cos\angle(q_i,\Delta q_i)|$. For association, we use Pearson correlation
between training frequency and final item norm. For serving, after masking a
user's training history, we report the fraction of the top 100 occupied by the
most frequent 1\% of training items. At its fully normalized endpoint, the
test-time method of Kim et al. induces the same per-user ranking as our
unit-item score \cite{kim2023testtime}. We compare this endpoint with raw inner
product inside one trained model to diagnose the contribution admitted by item
magnitude, not to introduce a mitigation method. This comparison is not an
endorsement of a train/serve mismatch and does not label either ranking
distribution as ideal.

\subsection{Testing the sufficient conditions}

Experiments~0--7 use cosine InfoNCE with in-batch negatives
\cite{oord2018cpc}, plain SGD unless stated otherwise, and a lookup item tower.
Table~\ref{tab:premise-ablation} varies one optimizer, objective, or encoder
property at a time. Configurations satisfying all sufficient conditions are
tangent at every observed update.
Momentum~$0.9$, Adam, a dot-product objective, a shared two-layer map, and a
linear map on nonorthogonal features produce non-tangent trajectories in these
tests. These violations demonstrate loss of the guarantee in concrete cases;
they do not establish that each condition is necessary for every trajectory.
For the shared two-layer and nonorthogonal-feature cases, only about 20\% of
observed angles fall within $90^\circ\pm10^\circ$, and the trajectories span
the full half-circle rather than concentrating around tangency.

\begin{table}[t]
\caption{Controlled tests of the sufficient conditions. ``Tangent'' refers to
realized output motion, not merely the output gradient.}
\label{tab:premise-ablation}
\centering
\scriptsize
\begin{tabular}{@{}cllc@{}}
\toprule
Exp. & Encoder diagnosed & Objective/update & Tangent?\\
\midrule
0 & embedding & cosine; momentum/Adam & no\\
1 & embedding & cosine; SGD & yes\\
2 & embedding & dot product; SGD & no\\
3 & deep user / item lookup & cosine; Adam/SGD & no/yes\\
4 & embedding + linear & cosine; SGD & no\\
5 & fixed features + linear & cosine; SGD & no\\
6 & one-hot + linear & cosine; SGD & yes\\
7 & embedding & $(1-\cos)^2$; SGD & yes\\
\bottomrule
\end{tabular}
\end{table}

Experiments~8 and~9 use two 100-dimensional embedding towers, the only difference is bigger and smaller learning rate, respectively. Full protocol is in \href{https://gitlab.com/wrapper228/usrw_pb/-/blob/main/README.md}{repository (click here)}.
Their contrast probes cumulative displacement, not a portable
nominal learning-rate threshold. Experiment~10 uses a BERT-like sequential user
tower \cite{sun2019bert4rec} with Adam and weight decay, and a disjoint
100-dimensional item lookup with SGD and a cyclic schedule
\cite{smith2017clr}; only the item side lies in the exact regime.

\begin{table*}[t]
\caption{Exact-regime outcomes. Paired entries are
MovieLens~32M/Amazon Books. Head share is the top-1\%-frequency fraction of the
top 100. Dot--unit differences diagnose item magnitude within a model.}
\label{tab:exact-results}
\centering
\small
\begin{tabular}{@{}clcccc@{}}
\toprule
Exp. & Request-side tower & Tangent item path & Frequency--norm $r$
& Top-1\% dot & Top-1\% unit item\\
\midrule
8 & embedding; larger cumulative motion
  & yes & $.66/.63$ & $.72/.60$ & $.29/.06$\\
9 & embedding; small cumulative motion
  & yes & $.0055/.0002$ & $.007/.010$ & $.007/.010$\\
10 & BERT-like; Adam + weight decay
   & yes & $.57/.61$ & $.63/.98$ & $.32/.11$\\
\bottomrule
\end{tabular}
\end{table*}

\begin{figure*}[t]
\centering
\includegraphics[width=.82\textwidth]{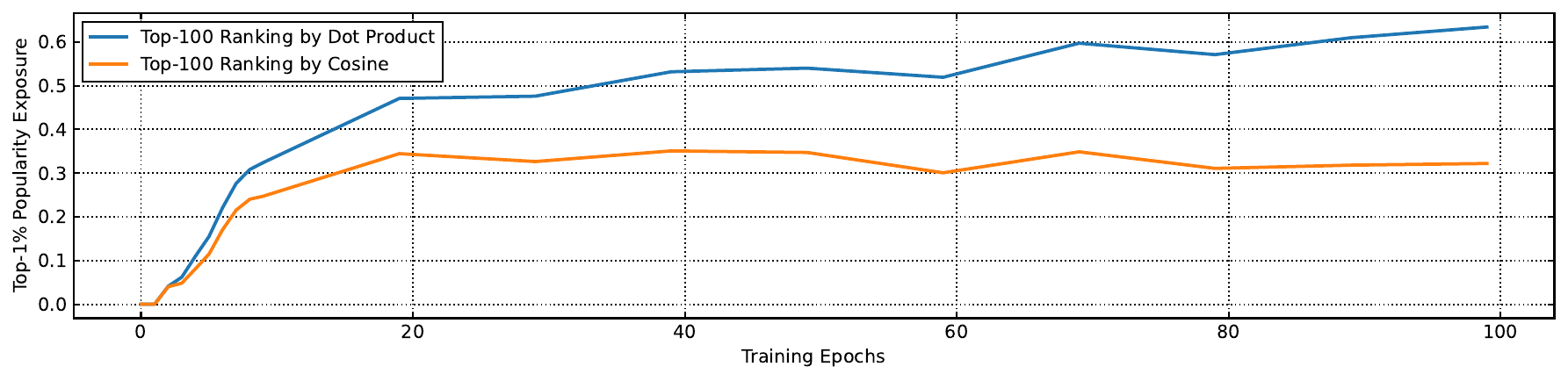}
\caption{Experiment~10 on MovieLens~32M. Item norms and raw-dot head share
increase as tangent item updates accumulate, while unit-item ranking excludes
magnitude. The trajectories visualize association, not causal mediation of all
head exposure by norm.}
\label{fig:exact-dynamics}
\end{figure*}

Tangency holds in both Experiments~8 and~9, but only the larger accumulated
path overcomes initialization-scale norm variation. Frequency--norm
correlations in Experiments~8 and~10 reject a zero-correlation null at
$p<.05$ on both datasets; Experiment~9 remains negligible. Thus exact tangency
is sufficient for per-step norm increase but not for a detectable
cross-item correlation. When correlation emerges, magnitude can substantially
raise head exposure under raw-dot ranking. The unit-item column may itself
contain relevance, log-exposure, or other popularity effects; only the
within-model gap belongs to the magnitude channel diagnosed here. These runs
are mechanism stress tests, not quality-optimized recommenders.

\subsection{A nonlinear tower beyond the guarantee}
\label{sec:hybrid}

We next study
\begin{equation}
 q_i=e_i+\alpha f_\theta(x_i),
 \label{eq:hybrid-output-main}
\end{equation}
where the ID row $e_i$ uses no-weight-decay Exact RowWise Adagrad
\cite{ivchenko2022torchrec,torchrec_tutorial}, and a shared two-layer GELU
content encoder uses AdamW with weight decay. The complete tower is nonlinear
and shares parameters across items. For an active item, let $d_{i,t}$ denote
the component of its output change caused by updating its ID row. Exact RowWise
Adagrad rescales the whole row gradient by one scalar, so this component is
exactly tangent to the current total item vector. Writing the simultaneous
change caused by the shared content encoder as $r_{i,t}$ gives
$\Delta q_{i,t}=d_{i,t}+r_{i,t}$ and
\begin{equation}
 \left|\cos\angle(q_{i,t},\Delta q_{i,t})\right|
 \le \frac{\|r_{i,t}\|}{\|\Delta q_{i,t}\|}.
 \label{eq:hybrid-bound}
\end{equation}
Tangency of the total item-vector movement is now empirical, but any deviation
has a localized source.

\paragraph{Protocol.}
We use global temporal 80/20 splits, fit user/item 3-cores and all popularity
statistics on training only, mask train-seen items, and evaluate warm users
with at least three future positives. MovieLens content contains genres and
release-decade buckets; Amazon uses hashed category, brand, and date metadata
with explicit missing values. A 256-dimensional hidden layer maps content into
the 128-dimensional item space; users are 128-dimensional ID vectors. Full 
protocol is in \href{https://gitlab.com/wrapper228/usrw_pb/-/blob/main/README.md}{repository (click here)}.

For each item, define total squared-norm change
$G_i=\|q_{i,T}\|^2-\|q_{i,0}\|^2$, accumulated tangent energy
$D_i=\sum_t\|d_{i,t}\|^2$, and signed residual $H_i=G_i-D_i$. The identity
$G_i=D_i+H_i$ is exact pathwise. With $z_i=\log(1+f_i)$, define
$C_X=\operatorname{Cov}_i(z_i,X_i)$. The signed ratio $C_D/C_G$ localizes a
frequency-associated term but is not a causal percentage.

\begin{table}[t]
\caption{Hybrid mechanism diagnostics (means across three seeds; selected
sample SDs shown). Arrowed $r$ is
initial$\rightarrow$final$\rightarrow$retracted frequency--norm correlation.}
\label{tab:hybrid-mechanism}
\centering
\scriptsize
\begin{tabular}{@{}lcc@{}}
\toprule
Metric & ML-32M & Amazon\\
\midrule
Active $|\cos|$ median/q95
 & $.0034/.0426$ & $.0003/.0095$\\
$\|r\|/\|\Delta q\|$ median/q95
 & $.0166/.0972$ & $.0019/.0384$\\
Frequency--norm $r$
 & $-.002\!\to\!.465\!\to\!.199$ & $-.000\!\to\!.404\!\to\!.148$\\
$C_D/C_G$
 & $1.104\!\pm\!.002$ & $.686\!\pm\!.001$\\
$r(z,N)$
 & $.569$ & $.439$\\
$r(z,D/N)$
 & $-.032$ & $-.629$\\
Content function change $M_f$
 & $2.420\!\pm\!.087$ & $1.467\!\pm\!.072$\\
\bottomrule
\end{tabular}
\end{table}

The angles of total active item-vector movements are much closer to tangency
than an isotropic 128-dimensional reference (median/q95 $=.0599/.1729$). The
content function is not approximately frozen: removing it worsens held-out loss in every seed,
although fine-tuning it over its frozen initialization helps only on Amazon.
The exact covariance closure holds in every run. On MovieLens, the residual
opposes the positive tangent covariance; on Amazon it reinforces it, while the
tangent term remains larger.

Most directly for Equation~\eqref{eq:two-stream-accumulation},
log-frequency correlates with ID update count but not positively with tangent
energy per update (Table~\ref{tab:hybrid-mechanism}). Accumulation therefore
follows additional opportunities, even though frequent Amazon items receive
smaller tangent energy per update.

\begin{table*}[t]
\caption{Hybrid serving and utility, means over three seeds. Head values are
fractions; lifts and reductions are percentage points. Recall/NDCG show
standard$\rightarrow$matched-retracted raw-dot runs.}
\label{tab:hybrid-serving-utility}
\centering
\scriptsize
\begin{tabular}{@{}lcccccc@{}}
\toprule
Dataset & Unit/actual head & Tangent-only lift & Standard/retracted norm lift
& Paired reduction & Recall@100 & NDCG@100\\
\midrule
ML-32M & $.495/.511$ & $9.87$ & $1.62/-11.98$
& $13.60\!\pm\!.88$ & $.1991\!\to\!.1644$ & $.2173\!\to\!.1768$\\
Amazon & $.692/.893$ & $18.96$ & $20.12/-.33$
& $20.45\!\pm\!.60$ & $.0398\!\to\!.0320$ & $.0170\!\to\!.0133$\\
\bottomrule
\end{tabular}
\end{table*}

The tangent-only counterfactual and learned norms raise head exposure on both
datasets, but standard norm lift is dataset-dependent: $1.62$ points on
MovieLens and $20.12$ on Amazon. Retracting each ID-row update's immediate norm
change lowers frequency--norm correlation in all six runs and shifts norm lift
downward; paired user-bootstrap 95\% intervals for the lift reduction exclude
zero in every run. Raw-dot Recall and NDCG also decrease in every run, by
roughly 17--22\% relative, with paired intervals excluding zero. Unit-item
utility changes are much smaller and mostly inconclusive. Retraction therefore
tests sensitivity to immediate norm growth caused by ID-row updates; it is
neither a norm-only
post-hoc edit nor a cost-free mitigation.

\section{Unified-Discovery Audit and Limits}
\label{sec:unified-implications}

Equation~\eqref{eq:two-stream-accumulation} makes the unified-system implication
specific. If search and recommendation update one direction-preserving item
row, their tangent energies accumulate in one norm. If either surface then
uses a shared raw-inner-product index or passes magnitude to a downstream
scorer, state learned from one stream can affect candidates on the other.
Neither joint training nor a shared catalog is sufficient by itself: the
relevant item parameters must be shared, the update path must retain a tangent
component, and serving must admit magnitude.

This motivates four task-separated checks:
\begin{enumerate}
  \item \textbf{Measure realized motion.} Log
  $q_i^\top\Delta q_i/(\|q_i\|\|\Delta q_i\|)$ after optimizer steps. Logging
  $q_i^\top\nabla_{q_i}\mathcal L$ alone tests only output-gradient geometry.
  \item \textbf{Attribute opportunities and energy.} Maintain
  $N_i^{\mathrm S},N_i^{\mathrm R}$ and accumulated
  $D_i^{\mathrm S},D_i^{\mathrm R}$, then report count and energy per event by
  task and frequency bucket. This distinguishes more opportunities from larger
  updates.
  \item \textbf{Decompose serving.} Within each trained model and surface,
  compare raw inner product with unit-item scoring after identical candidate
  masking. Their difference diagnoses magnitude; it does not define unbiased
  relevance.
  \item \textbf{Test cross-task sensitivity.} Compare search-only,
  recommendation -only, and joint updates of the shared item tower, followed by
  both serving surfaces. A matched retraction can test trajectory sensitivity,
  but its utility must be reported and it should not be described as pure
  mediation.
\end{enumerate}

\paragraph{What transfers mathematically.}
The tangent-update and additive-accounting statements are agnostic to whether
a realized event came from an explicit query or an implicit recommendation
context. Consequently, a lookup-equivalent shared row satisfying the theorem
has the same per-update norm law in either stream. This is the supported
cross-task result.

\paragraph{What remains empirical.}
Our datasets contain recommendation interactions, not paired search and
recommendation logs such as KuaiSAR \cite{sun2023kuaisar}. Query traffic can
change item frequencies, negative construction, relevance supervision,
request directions, and score calibration. We therefore do not know whether
the tangent path is prevalent or material in a deployed joint model, whether
search or recommendation contributes more energy, or whether the induced
magnitude improves either task. The proposed task-separated experiment is
future validation, not evidence reported here.

\paragraph{Other limitations.}
The exact theorem is sufficient and covers a narrow direction-preserving row
path; it is not a characterization of AdamW-trained shared MLPs. The hybrid
experiment establishes existence beyond the exact regime for one construction,
not prevalence across architectures. Frequency is observational in the data,
and $G=D+H$ is an accounting identity rather than a causal mediation estimator.
The retraction changes subsequent directions and lowers raw-dot utility, so it
does not identify a free debiasing strategy. Finally, head-item exposure can
encode useful priors as well as unwanted amplification; application-specific
relevance, fairness, and business objectives are required to judge it.

\section{Conclusion}

Cosine-gradient orthogonality is a property of an encoder output; tangent
representation motion is a property of the parameter update. A
lookup-equivalent item path with direction-preserving row updates bridges the
two, making every step add its squared length to item squared norm. When search
and recommendation share that path, their update opportunities conditionally
accumulate in the same magnitude, which raw-inner-product serving can expose on
both surfaces.

Condition tests validate the predicted exact behavior, while a nonlinear tower
that combines per-item ID vectors with a shared content encoder shows that a
traceable tangent component can remain material outside the guarantee. Its
serving effect varies sharply by dataset,
and suppressing immediate norm growth reduces utility. The practical result is
therefore an audit, not a universal bias or mitigation claim: measure realized
motion, separate task-specific opportunity count from energy per update, and
decompose direction from magnitude at serving. Joint search--recommendation
logs are the necessary next test of cross-surface prevalence.

\bibliographystyle{ACM-Reference-Format}
\bibliography{references}
\end{document}